\newif\ifjcp
\jcptrue  

\ifjcp
    \documentclass[twocolumn,aip,jcp,reprint,floatfix]{revtex4-2}
\else
    \documentclass[journal=jacsat,manuscript=article]{achemso}
    \setkeys{acs}{articletitle = true}
    
    \SectionNumbersOn
\fi

\usepackage{float}
\usepackage{xcolor}
\usepackage{graphicx}
\usepackage{dcolumn}
\usepackage{bm}
\usepackage{amsmath}
\usepackage{braket}
\usepackage{dsfont} 
\usepackage{hyperref}

\usepackage{balance}

\usepackage{fancyhdr}
\usepackage{fnpos}
\usepackage[english]{babel}

\usepackage{array}

\usepackage{nicefrac}
\usepackage{amssymb}

\DeclareMathOperator*{\argmax}{argmax} 

\newcommand{\toremove}[1]{\ignorespaces}

\newcommand{\setinfo}{%
    \title{Fast and accurate nonadiabatic molecular dynamics enabled through variational interpolation of correlated electron wavefunctions}%
    \author{Kemal Atalar}%
    \email{kemal.atalar@kcl.ac.uk}
    \affiliation{Department of Physics and Thomas Young Centre, King’s College London, Strand, London, WC2R 2LS, UK}%
    \author{Yannic Rath}%
    \affiliation{Department of Physics and Thomas Young Centre, King’s College London, Strand, London, WC2R 2LS, UK}%
    \affiliation{National Physical Laboratory, Teddington, TW11 0LW, UK.}%
    \author{Rachel Crespo-Otero}%
    \affiliation{Department of Chemistry University College London, 2020 Gordon St., London, WC1H 0AJ, UK.}%
    \author{George H. Booth}%
    \email{george.booth@kcl.ac.uk}%
    \affiliation{Department of Physics and Thomas Young Centre, King’s College London, Strand, London, WC2R 2LS, UK}%
}

\ifjcp
\else
\fi

\ifjcp\else
    \setinfo
\fi
\begin{document}
\ifjcp
    \setinfo
\fi
\newcommand*{\ka}[1]{\textcolor{blue}{KA: #1}}
\newcommand*{\ghb}[1]{\textcolor{cyan}{George: #1}}
\newcommand{\ubar}[1]{\text{\b{$#1$}}}

\begin{abstract}
We build on the concept of eigenvector continuation to develop an efficient multi-state method for the rigorous and smooth interpolation of a small training set of many-body wavefunctions through chemical space at mean-field cost. The inferred states are represented as variationally optimal linear combinations of the training states transferred between the many-body basis of different nuclear geometries. We show that analytic multi-state forces and nonadiabatic couplings from the model enable application to nonadiabatic molecular dynamics, developing an active learning scheme to ensure a compact and systematically improvable training set. This culminates in application to the nonadiabatic molecular dynamics of a photoexcited 28-atom hydrogen chain, with surprising complexity in the resulting nuclear motion. With just 22 DMRG calculations of training states from the low-energy correlated electronic structure at different geometries, we infer the multi-state energies, forces and nonadiabatic coupling vectors at 12,000 geometries with provable convergence to high accuracy along an ensemble of molecular trajectories, which would not be feasible with a brute force approach. This opens up a route to bridge the timescales between accurate single-point correlated electronic structure methods and timescales of relevance for photo-induced molecular dynamics.
\end{abstract}

\ifjcp
    \maketitle
\fi



\section{Introduction}

Nonadiabatic molecular dynamics (NAMD) is essential for an {\em ab initio} simulation of chemical processes where nuclei move over reaction pathways involving the participation of multiple electronic states~\cite{Curchod2018-aiNAMDreview,CrespoOtero18-chemicalreview,Nelson2020-NAMDreview,Matsika2021-electronicstructureCoIns}. These states are commonly accessed via photo-excitation, but catalytic, thermal and other reactive processes can also involve coupling excited electronic states and the nuclear motion beyond the adiabatic regime~\cite{Nelson2020-NAMDreview}. These processes underlie internal conversion and intersystem crossings, whose rates critically depend on accurate energetic gaps between the energy levels and their nonadiabatic coupling strengths at the different nuclear geometries. The presence and location of conical intersections is key to modelling these nonadiabatic processes~\cite{Matsika11-CoInReview}, and an outstanding challenge in computational chemistry in this important field.

While there are a number of alternative formulations for efficient propagation of coupled electron and nuclear motion, for many purposes  mixed quantum-classical approaches can be devised whereby the nuclear propagation is still treated classically, relying on information from all relevant electronic states in the dynamics. These electronic states can therefore be obtained under the Born-Oppenheimer (BO) approximation~\cite{BornOppenheimer} at each nuclear geometry without explicit coupling to quantum nuclear degrees of freedom. In this work, we consider the popular `fewest-switches surface hopping' (FSSH) approach~\cite{Tully90-FSSH,Subotnik16-NAMDreview_withdecoherence,Jan15-fsshdecoherence}, although many other approaches also exist~\cite{CrespoOtero18-chemicalreview,Curchod2018-aiNAMDreview}.
In FSSH, the full electronic state is evolved at each timestep as a superposition over multiple Born-Oppenheimer adiabatic electronic states, with the nuclei propagated on a single adiabatic electronic surface. When the simulations are done using adiabatic states, nonadiabatic stochastic jumps between the electronic states 
occur with a probability depending on the nonadiabatic couplings, whilst ensuring overall energy conservation. While this approach can be motivated from higher levels of theory\cite{Subotnik13-FSSHfromLiouville}, it lacks rigorous interference and decoherence effects of quantum nuclei (noting recent work incorporating nuclear decoherence and tunneling processes into the formulation~\cite{Granucci07-decoherence,Granucci10-decoherence,Subotnik16-NAMDreview_withdecoherence,Jan15-fsshdecoherence,BenNun02-tunnelingNAMD}). Provided \textit{ad-hoc} corrections are incorporated to address some of these issues, FSSH has proven to be very effective in the modelling and prediction of photo-chemical processes over relevant atomic timescales\cite{CrespoOtero18-chemicalreview}. 

However, the outstanding limitation in the scope of application of NAMD is the electronic structure problem at its heart\cite{Matsika2021-electronicstructureCoIns,Plasser2014-singlereferenceNAMD},  which has found to be generally more important to the quality of results than the specifics of different NAMD approximations~\cite{Janos2023-electronicvsdynamics}. This is a demanding electronic structure challenge, requiring many consecutive calculations as the nuclei are propagated in time, each with a finely balanced description of multiple electronic states. Furthermore, nuclear forces are required for each state to determine the propagation of the nuclei, as well as nonadiabatic couplings (NACs) between the states for the propagation of the electronic part and stochastic hopping. Although it is possible in some cases to approximate these gradients and NACs if they are not available analytically from the electronic structure solver used, this introduces additional approximations, uncertainty and potentially overheads for the calculation. While density functional theory is almost ubiquitous for {\em ab initio} molecular dynamics on ground states, the requirement of multiple electronic states limits its applicability to NAMD (noting recent approximations for DFT-based NAMD, especially in simulations of transport\cite{Barbatti2016,doi:10.1021/ct400641n}). Furthermore, single-reference electronic structure theory based on linear response (such as TD-DFT\cite{TDDFT-NAMD}, CC2\cite{CHRISTIANSEN95-CC2} or EOM-CCSD\cite{Stanton93-EOMCCSD}) are also generally not suitable where transitions to the ground state are required, due to their explicit formulation as excitations from a single electronic state, which precludes the critical region around conical intersections where this gap vanishes and ground and excited states must be treated on the same footing~\cite{Plasser2014-singlereferenceNAMD}. These descriptions may not give the required balance in the accuracy of the ground and excited states, especially where excited states have significant charge rearrangement compared to the ground state.

This places a significant burden on multireference electronic structure methods for the description of these adiabatic states, in particular state-averaged CASSCF which can be considered the workhorse of NAMD, and which has analytic gradients and NACs\cite{Lengsfield84-NAC_CASSCF,Fdez16-NAC_CASSCF} available in many packages. Unfortunately, there are several drawbacks in these traditional multireference approaches, in particular the lack of dynamical correlations, which can unbalance the description of different states and manifest in e.g. dramatic overestimation of the relative energy of ionic states\cite{Slavicek10-casscf_imbalance,Angeli09-casscf_overestimate}. Extensions to internally-contracted MRPT and MRCI methods are more recently available, but they can still underestimate vertical excitations~\cite{Bomble04-caspt2_verticalexcitation} and are substantially more expensive with more challenging gradient theory and NACs~\cite{Park2017-CASPT2_NAC,Park2017-CASPT2_NAMD}. More fundamentally, these approaches depend significantly on the choice of active space\cite{Szymczak11-NAMD_CASdependence} and require the definition of a small active space to enable them to be tractable, which must remain consistent across the changes in geometry over the trajectory. This can often be hard or impossible, as relevant orbitals at one geometry can adiabatically change into an irrelevant subspace at another geometry, while orbitals crossing and entering or leaving the subspace can result in a discontinuous surface,  difficulties with intruder states and issues with energy conservation during the NAMD simulations. \cite{Plasser2012-review,Iino23-DMRG_gradient,Park2020-multireferencePES} 

This context of NAMD for photochemistry, catalysis and beyond, is a prime motivation for developments in electronic structure methodology. However, new wavefunction-based approaches emerging in the last couple of decades have not yet impacted upon this field, underlining the challenges faced in this area. Modern accurate and systematically improvable approaches such as DMRG\cite{10.1063/1.5129672, Reiher20-DMRGexcited, Chan11-DMRGreview}, selected CI\cite{Renzo87-selectedCI,Holmes2016-selectedCI}, and a number of stochastic methods (including FCIQMC\cite{Booth09-FCIQMC,Kai20-FCIQMC,Halson20-FCIQMC}, AFQMC\cite{Motta18-AFQMC,Lee2022-AFQMC} and advances in VMC models\cite{Choo2020-VMC,Rath23-VMC}) can in principle be used as multiconfigurational solvers within NAMD, at least to extend the size of active spaces and mitigate the difficulty in their appropriate selection. However, despite much progress in accurately describing excited states within these frameworks \cite{Neuscamman23-excitedQCandMC,Baiardi2022-DMRGexcited,Reiher20-DMRGexcited,Blunt15-FCIQMCExcited,Lischka2018-multireference_excited}, as well as their analytic nuclear gradients \cite{Hu2015-DMRGgradients,Iino23-DMRG_gradient,Thomas15-FCIQMC_force,Jiang22-FCIQMC_force,Chen23-AFQMC_force}, the community generally still lack nonadiabatic couplings in these methods, while stochastic noise in the electronic structure can often be challenging for precise molecular dynamics that conserves total energy (noting significant work \cite{Sorello15-QMC_MD}). More generally, while these methods can be powerful in obtaining a small number of single-point energies (often both for ground and excited states), they are still expensive for the number of calculations required in molecular dynamics. Furthermore, they often lack a robust black-box simulation protocol that hinders the reliable execution of these approaches for several thousand consecutive calculations, each relying on consistent convergence for all prior points for reliable trajectories, without manual checking of convergence and simulation parameters. This puts the timescales necessary to describe realistic photochemical processes out of their reach (a fact often even true with a CASSCF description).

As an alternative paradigm, machine learning (ML) potentials have been tremendously successful in breaking this computational barrier for ground state potential energy surfaces (PES). However, the application of ML methodologies to NAMD has various limitations and challenges (beyond those well-understood in ground-state ML force fields such as the local energy decomposition) despite showing great promise by accessing nanosecond NAMD trajectories for small molecules both with MR-CISD\cite{Westermayr19-MLnanosecond} and CASSCF\cite{Lopez21-MLnanosecond} training data. Although progress has been made in predicting excited state energies and forces in these frameworks~\cite{Pronobis2018-MLexcited,Dral20-MLexcited,Dral2021-MLexcited,Lopez21-MLnanosecond,Westermayr19-MLnanosecond,Westermayr20-SchnetSHARC,Axelrod2022-MLNAMD}, direct learning of NACs tend to lead to their underestimation~\cite{Axelrod2022-MLNAMD,Lopez22-insideMLNAMD}, while approximations to them based purely on energies and gradients can also lead to a substantial misrepresentation of true nonadiabatic processes~\cite{Westermayr20-SchnetSHARC,Casal22-BaeckanNAMD,Merritt2023-openMOLCAS_NX_interface}. Capturing conical intersections (CoIns) is also a challenge, as small and sharp energy gaps near CoIns tend to be overestimated and smoothened since they are necessarily poorly represented in the training set~\cite{Westermayr20-MLperspective,Lopez22-insideMLNAMD}. Furthermore, all of the same electronic structure challenges still exist in obtaining appropriate training data required to build these models~\cite{Westermayr20-MLperspective}. 

The approach outlined in this work takes a step to address these shortcomings, in a largely method-agnostic framework for the robust and compact interpolation of accurate wave functions through chemical space. Initially described in Ref.~\cite{Rath24-evcont} for the interpolation of ground states, we show how this approach can be extended for the balanced interpolation of both ground and excited states for \emph{ab initio} systems over changing atomic geometries with mean-field cost. Importantly, we also show how both excited state gradients and nonadiabatic couplings between the states are straightforwardly extracted from the interpolation framework, allowing application to NAMD. This allows for the first (to the best of our knowledge) NAMD simulation using modern DMRG derived electronic states~\cite{10.1063/1.4905329,10.1063/5.0050902}, obtaining high-accuracy molecular dynamics of excited hydrogen chains -- a system for which we believe no other solver would be effective. We discuss both the convergence of the dynamics with respect to increasing number of training DMRG calculations supporting the interpolated model, and an approach for efficient selection of these training geometries. Finally, we end with a perspective of the use of this framework more generally for practical NAMD applications.


\section{Variational Wavefunction Interpolation}

We introduce here the `eigenvector continuation' approach that we use in this work to interpolate a representative set of `training' many-body wavefunctions through chemical space as the nuclei move. This is essentially a multi-state generalization of the approach introduced in Ref.~\cite{Rath24-evcont}, and we refer the reader to this for a fuller exposition. That approach was inspired by developments in the nuclear physics and lattice model communities~\cite{PhysRevLett.121.032501,PhysRevC.101.041302,duguet2023eigenvector,KONIG2020135814,PhysRevC.107.064316,DRISCHLER2021136777,10.1063/5.0141145}, and a highly-related approach was also developed for interpolating {\em ab initio} systems with quantum computers~\cite{mejutozaera2023-evcont}. The method can be motivated from either a machine-learning perspective or as a subspace projection method, and we take the latter approach here. 

We assume that we have a `training' set of $M$ non-orthogonal many-body states, as vectors in the electronic Hilbert space, $\{ |C_a\rangle \} = C_{\mathbf{n}}^{(a)}$, where $a,b,\dots$ labels these distinct training states and $\mathbf{n}$ denote the occupation number vectors of the orthonormal many-electron configurations at the nuclear geometry of interest, $\mathbf{R}$. We project the {\em ab initio} Hamiltonian at this nuclear configuration into the space spanned by these states. We can then consider a valid wave function at this geometry as resulting from a variational optimization within the span of this many-body subspace. This can be found in closed form as a generalized eigendecomposition of the $M \times M$ Hamiltonian in this basis,
\begin{equation}
    \langle C_a | \mathcal{H}(\mathbf{R}) | C_b \rangle \, x_{b}^{(A)} (\mathbf{R}) = E_{A} (\mathbf{R}) \, \langle C_a | C_b \rangle \, x_{b}^{(A)} (\mathbf{R}) . \label{eq:subspaceH}
\end{equation}
The training basis (comprising $M$ elements) is small enough that this generalized eigenvalue problem can be completely solved at all geometries of interest, giving a variational approximation to both the ground state and excitation spectrum at each arbitrary `test' geometry, $E_A(\mathbf{R})$. The eigenvectors, $x_{b}^{(A)}(\mathbf{R})$, denote the component of the many-body training vector in each interpolated state (these interpolated states are denoted by upper-case indices $A,B,\dots$). Through this scheme, wavefunctions at arbitrary test geometries and their observables can be inferred by sampling few wavefunctions at training geometries. As this training subspace is enlarged, the energies of all states from the subspace must necessarily variationally lower towards their exact eigenvalues, as ensured by the eigenvalue interlacing theorem~\cite{interlacingtheorem}. Due to the linearity of the model, the number of electronic states described remains constant as geometries change, and their energies must vary smoothly with changes in the nuclear potential (away from state crossings). The key questions now are how these training states are chosen such that they faithfully span the low-energy eigenstates as geometries change, as well as how the subspace Hamiltonian can be efficiently constructed without requiring the training data expressed in the exponentially large many-electron Hilbert space.

As indicated via the explicit dependence on $\mathbf{R}$ for certain quantities in Eq.~\ref{eq:subspaceH}, the training vectors are defined such that their numerical values remain fixed in an abstract orthonormal Hilbert space, regardless of the test geometry $\mathbf{R}$ at which we are evaluating the subspace model. This critically ensures that the overlap metric between many-body training states, $\langle C_a | C_b \rangle$, is independent of $\mathbf{R}$. However, if this training basis is to be interpreted as a set of physical wave functions at each geometry rather than abstract vectors, then it is worth stressing that their character does indeed change with $\mathbf{R}$, since the underlying Hilbert space of electronic configurations will change. 
Therefore, it is essential that we have a consistent and orthonormal representation of the Hilbert space at each geometry, such that the numerically fixed training vectors are transferrable between geometries and still span a space of relevance for the low-energy states of interest, rather than them spanning increasingly irrelevant parts of the many-body Hilbert space. 

To motivate a judicious choice for these training vectors, we assume that they come from the exact (FCI) solution for a small number of low-energy eigenstates of the Born--Oppenheimer electronic Hamiltonian at select `training' geometries of the system. However, we also need to fix a consistent representation of the orbitals, $\chi_i(\mathbf{r};\mathbf{R})$, which define the many-body Hilbert space for these vectors, such that the probability amplitudes of states $\{|C_a\rangle \}$ can be effectively transferred between geometries. To this end, we choose the orbital basis of the training states, $\chi_i(\mathbf{r};\mathbf{R})$, to be the symmetrically (Löwdin) orthonormalized atomic-orbital (SAO) basis~\cite{10.1063/1.1747632, https://doi.org/10.1002/qua.981}. These are simply and uniquely derived from an underlying atom-centered atomic orbital basis, $\{\phi_\alpha(\mathbf{r};\mathbf{R})\}$, as
        \begin{equation}
        \chi_i(\mathbf{r};\mathbf{R}) = \sum_\alpha \, [\mathbf{S}(\mathbf{R})]^{-1/2}_{\alpha i} \, \phi_\alpha(\mathbf{r};\mathbf{R}).
        \label{eq:orbital_def}
    \end{equation}
    where $\mathbf{S}(\mathbf{R})$ is the atomic orbital overlap matrix
    \begin{equation}
        S_{\alpha \beta}(\mathbf{R}) = \int d \mathbf{r} \, \phi^\ast_\alpha(\mathbf{r};\mathbf{R}) \, \phi_\beta(\mathbf{r};\mathbf{R}).
    \end{equation}
These SAOs are defined to remain (in a least-squares sense) as close as possible to the underlying local atomic orbital basis, while ensuring the required orthonormality of this abstract basis at each atomic geometry~\footnote{It is possible to work directly with a non-orthogonal AO representation of the abstract many-electron representation of the training states, but this then requires a rotation of the many-body states at each geometry. This was explored in Ref.~\cite{mejutozaera2023-evcont}, but entails exponential complexity for each inference and therefore it primarily motivated for application for quantum computers.}.
This choice of representation for the training states $\{|C_a\rangle \}$ (obtained over a range of geometries) is motivated by the fact that much of the local character of the correlated many-electron state in this basis will remain similar as atoms move by small amounts. In addition, nearby atoms with similar chemical bonding will also have common features in their many-electron quantum fluctuations characterizing e.g. covalent bonding character. Therefore, the numerical values of the probability amplitudes of the states in this representation will plausibly remain `close' to the states of interest at modified geometries, ensuring that the numerical values of the FCI many-electron states change least between the different electronic Hilbert spaces as the atomic positions change, and that the states can act as a general projector into the low-energy space as atoms move.

While this is a heuristic choice, we have the rigorous conditions that the inferred wave functions are strictly variational (for all geometries) with respect to increasing training data, as well as the exactness of all inferred states at test geometries which coincide with training geometries. From an ML perspective in its application to MD, the variationality of the model therefore ensures an inductive bias away from regions of the phase space which are poorly represented by the training data. The inferred states are at all points represented as a variationally optimal linear combination of the training states in their SAO representations, as
\begin{equation}
    \ket{\Psi_A(\mathbf{R})} = \sum_a \sum_\mathbf{n} x_a^{(A)^*}(\mathbf{R}) \, C_{\mathbf{n}}^{(a)} \ket{\mathbf{n}}.
    \label{eq:evcont_expansion}
\end{equation}
No special structure is relied upon for the description of any of these states (other than the fact that their SAO-represented FCI vectors over the geometries of interest remain sufficiently close to a linear combination of the training states). Indeed, no mean-field information is used at any point in the framework. This provides confidence that strong electronically correlated states can be described, and that the procedure should be relatively unbiased for a description of multiple electronic states simultaneously, providing those states are equally represented in the training states. Furthermore, in contrast to widespread machine-learning derived force fields now prevalent in molecular dynamics, no local energy decomposition is employed, and the fact that a valid many-electron state is propagated to all geometries means that all properties of interest can be predicted from the same model. This includes analytic nuclear forces, which are particularly straightforward due to the geometry-independence of the subspace definition and variational formulation~\cite{Rath24-evcont}. Furthermore, as we shall show in Sec.~\ref{sec:nac}, nonadiabatic coupling vectors between inferred states are also straightforwardly obtainable, which places the approach as a suitable candidate as an electronic structure solver for NAMD. Testing these assertions is at the heart of this work.

Finally, we note that FCI probability amplitudes in a local SAO representation are invariant to translation and rigid body rotations of the molecule (provided a consistent ordering of the underlying AOs). Furthermore, at dissociation the SAO probability amplitudes are unchanging for all states regardless of the extent of the dissociation, ensuring that all inferred wave functions in this important strongly-correlated limit should be consistently described. The appropriate choice of training geometries in which to support the model across e.g. a molecular dynamics trajectory is however critical to the success of the method. In Sec.~\ref{sec:active-learning} we develop an active-learning protocol to greedily update the training states in a self-consistent procedure across an MD trajectory, demonstrating convergence to near-exactness of NAMD over the trajectory.




\subsection{Density matrix formulation and training from DMRG}

For a practical implementation of the scheme outlined above, it is essential that the projected Hamiltonian of Eq.~\ref{eq:subspaceH} at each geometry can be constructed efficiently, without requiring manipulation of the underlying training states with their exponential complexity. The overlap of the training states, $\langle C_a | C_b \rangle=\sum_{\mathbf{n}} C_\mathbf{n}^{(a)*} C_\mathbf{n}^{(b)}$, is independent of geometry and therefore can be precomputed for the training states and reused at each inference point. 
Similarly, the {\em ab initio} Hamiltonian at any test point can be efficiently projected into the training space via the one- and two-body transition density matrices (tRDMs) between all pairs of training states, ($a$, $b$). These can be precomputed from the training data and are denoted $\gamma_{ab}^{ij}$ and $\Gamma_{ab}^{ijkl}$ respectively. We write the Hamiltonian as
\begin{equation}
    \mathcal{H}(\mathbf{R}) = \sum_{ij} h^{(1)}_{ij}(\mathbf{R}) \, \hat{c}^\dagger_{i} \hat{c}_{j} + \frac{1}{2} \sum_{ijkl} h^{(2)}_{ijkl}(\mathbf{R}) \, \hat{c}^\dagger_{i} \hat{c}^\dagger_{j} \hat{c}_{l} \hat{c}_{k} 
\label{eq:hamiltonian} 
\end{equation}
enabling the projection to be found as
\ifjcp
\begin{align}
    \langle C_a | \mathcal{H}|C_b\rangle (\mathbf{R})  &= 
     \sum_{ij} \sum_{\mathbf{n} \mathbf{n}'} C_{\mathbf{n}}^{(a)*} C_{\mathbf{n}'}^{(b)} \langle \mathbf{n} | {\hat c}^\dagger_i {\hat c}_j | \mathbf{n}' \rangle \, h_{ij}^{(1)}(\mathbf{R}) \nonumber \\
     &\quad + \frac{1}{2} \sum_{ijkl} \sum_{\mathbf{n} \mathbf{n}'} C_{\mathbf{n}}^{(a)*} C_{\mathbf{n}'}^{(b)} \langle \mathbf{n} | {\hat c}^\dagger_i {\hat c}^\dagger_j {\hat c}_l {\hat c}_k | \mathbf{n}' \rangle \, h_{ijkl}^{(2)}(\mathbf{R}) \nonumber \\
    &= 
    \sum_{ij} \gamma^{ij}_{ab} \, h_{ij}^{(1)}(\mathbf{R}) +
    \frac{1}{2} \sum_{ijkl} \Gamma^{ijkl}_{ab} \, h_{ijkl}^{(2)}(\mathbf{R}), \label{eq:subspaceh}
\end{align}
\else 
\begin{align}
    \langle C_a | \mathcal{H}|C_b\rangle (\mathbf{R})  &= 
     \sum_{ij} \sum_{\mathbf{n} \mathbf{n}'} C_{\mathbf{n}}^{(a)*} C_{\mathbf{n}'}^{(b)} \langle \mathbf{n} | {\hat c}^\dagger_i {\hat c}_j | \mathbf{n}' \rangle \, h_{ij}^{(1)}(\mathbf{R}) + 
     \frac{1}{2} \sum_{ijkl} \sum_{\mathbf{n} \mathbf{n}'} C_{\mathbf{n}}^{(a)*} C_{\mathbf{n}'}^{(b)} \langle \mathbf{n} | {\hat c}^\dagger_i {\hat c}^\dagger_j {\hat c}_l {\hat c}_k | \mathbf{n}' \rangle \, h_{ijkl}^{(2)}(\mathbf{R}) \nonumber \\
    &= 
    \sum_{ij} \gamma^{ij}_{ab} \, h_{ij}^{(1)}(\mathbf{R}) +
    \frac{1}{2} \sum_{ijkl} \Gamma^{ijkl}_{ab} \, h_{ijkl}^{(2)}(\mathbf{R}), \label{eq:subspaceh}
\end{align}
\fi
Crucially, once these tRDMs and overlaps between the training states are known, the simulation proceeds with the subspace Hamiltonian constructed via the contraction of Eq.~\ref{eq:subspaceh} in $\mathcal{O}[M^2 L^4]$ cost, where $M$ is the number of training states and $L$ is the number of basis functions in the system. This relatively low polynomial scaling contrasts with the generally exponential costs of accurate wave function solutions to the electronic structure problem to compute the training states, highlighting the significant speed up in this interpolation when many calculations are required across chemical space (and especially where multiple electronic states are required). Properties can then be extracted from the inferred state via its one- and two-body reduced density matrices (RDMs) represented in the SAO basis of each geometry, without any explicit recourse to the training states, as
\begin{equation}
    \gamma^{ij}_{(A)} = \sum_{a,b}^M x_a^{(A)^*} \, \gamma^{ij}_{ab} \, x_b^{(A)},
\end{equation}
and
\begin{equation}
    \Gamma^{ijkl}_{(A)} = \sum_{a,b}^M x_a^{(A)^*} \, \Gamma^{ijkl}_{ab} \, x_b^{(A)}.
\end{equation}

\begin{figure*}
\centering
\includegraphics[width=\textwidth]{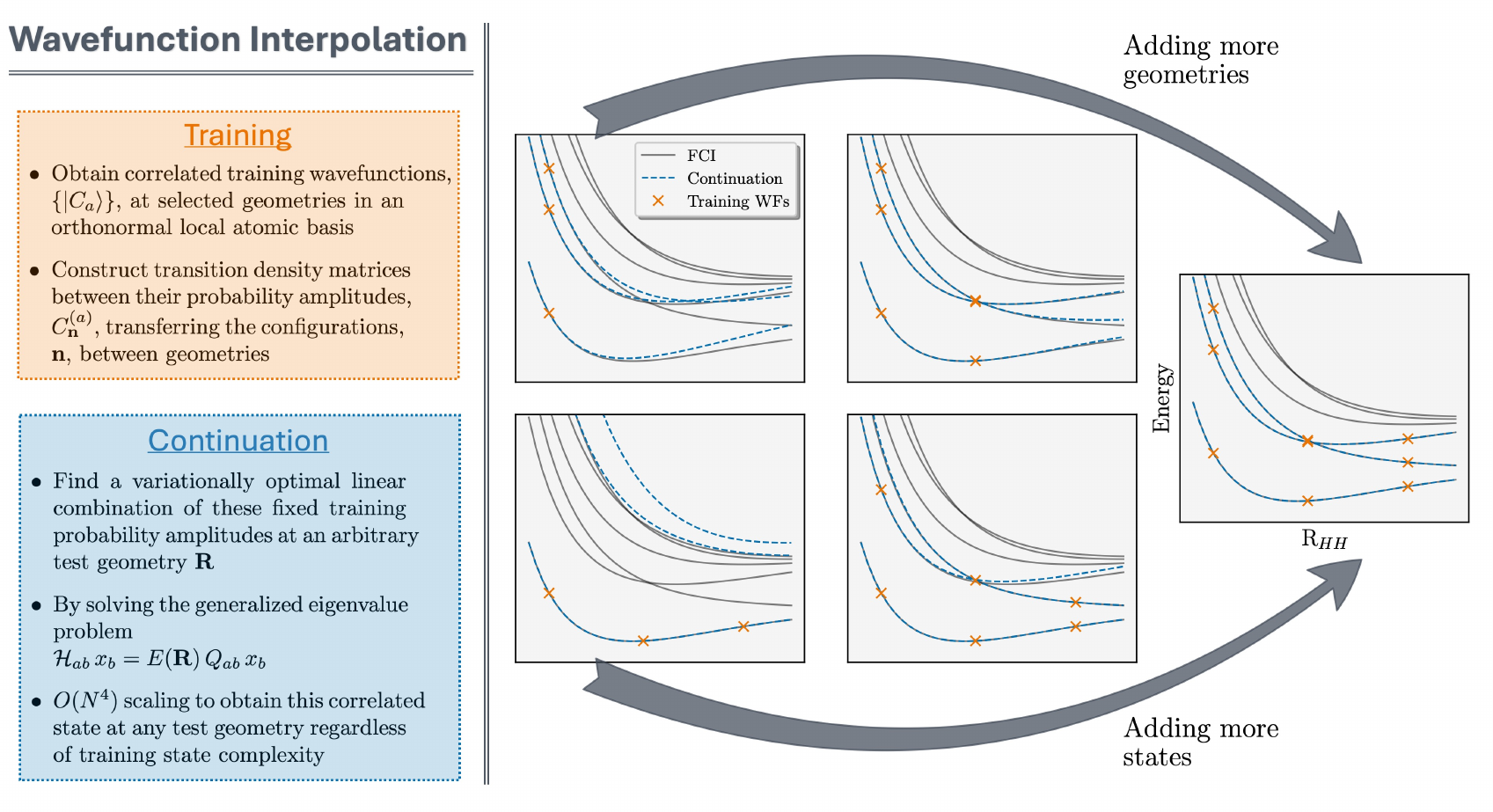}
  \caption{Demonstration of the eigenvector continuation scheme for interpolating the lowest three states of a symmetrically stretched linear four-atom hydrogen chain in an STO-3G basis compared to FCI results (black). Orange crosses represent the training wave functions used for each panel.}
  \label{fig:evcont-demo}
\end{figure*}

This framework is demonstrated in Fig.~\ref{fig:evcont-demo}, where we show a simple proof-of-principle for the one-dimensional phase space corresponding to the symmetric stretch of four Hydrogen atoms. The six lowest-energy FCI solutions are shown, including a state-crossing, and the training states supporting the interpolation in each panel indicated by orange crosses. For the interpolated state, the span of the training space can be enlarged systematically by including either a larger number of higher-lying eigenstates at each geometry (which will necessarily be orthogonal at the same geometry), or a larger number of geometries (which will be non-orthogonal between different geometries). We show that the three lowest lying states can be smoothly interpolated to near-exactness (including a state crossing) with just three geometries, each contributing three states to the training space. This system is further benchmarked in Fig.~\ref{fig:NAC}. In all results of this work, we select the same number of training states at each geometry as the number of low-energy states of interest for the interpolation unless stated otherwise.

Finally, we note that while the framework was described with the use of exact (FCI) training states, other approximate electronic structure methods could also be used to define these training states.
As long as $N$-representable tRDMs corresponding to physical wave functions, as well as overlaps, are defined in the method, then any approach could be used within the scheme and a variational state would be inferred at all points, motivating the framework as a tool to accelerate many different electronic structure methods. For the larger systems in Sec.~\ref{sec:largesys}, we therefore use the density matrix renormalization group (DMRG)\cite{PhysRevLett.69.2863}, which defines training states as matrix product states.


\subsection{Ground and excited state forces}
\label{sec:forces}

The propagation of nuclear trajectories in molecular dynamics relies on an accurate and efficient evaluation of nuclear forces. For nonadiabatic MD, this also requires energy derivatives for the excited state potential energy surfaces. This follows a similar derivation to the ground state energy gradients which were introduced within the framework of the interpolation scheme in Rath et al.~\cite{Rath24-evcont}, and which can be readily generalized to excited states. We partition the force into nuclear and electronic contributions as
\begin{equation}
    \mathbf{F}_A(\mathbf{R}) = - \frac{\partial E_\mathrm{tot}^{(A)}}{\partial \mathbf{R}} = - \Big( \frac{\partial E_\mathrm{nuc}}{\partial \mathbf{R}} + \frac{\partial E_A}{\partial \mathbf{R}} \Big),
\end{equation}
where $E_\mathrm{nuc}$ is the nuclear contribution to the energy. 
The variationality of the inferred state along with the $\mathbf{R}$-independent subspace projectors allow a state-specific electronic contribution to the force to be written using the Hellman-Feynman theorem\cite{PhysRev.56.340} as
\begin{equation}
     \frac{\partial E_A}{\partial \mathbf{R}} =  
    \frac{\partial }{\partial \mathbf{R}} \Big(\mathbf{x}^{(A)^\dagger} \, \mathcal{H} \, \mathbf{x}^{(A)} \Big) = \mathbf{x}^{(A)^\dagger} \, \frac{\partial \mathcal{H}}{\partial \mathbf{R}} \, \mathbf{x}^{(A)},
    \label{eq:dE_dR_first}
\end{equation}
where $\mathbf{x}^{(A)}$ is the eigenvector of Eq.~\ref{eq:subspaceH} corresponding to the $A$th inferred state.
Substituting in the Hamiltonian of Eq.~\ref{eq:hamiltonian}, the electronic contribution to the force can be evaluated for any state $A$ as
\begin{equation}
    \frac{\partial E_A}{\partial \mathbf{R}} = \sum_{ij} \gamma^{ij}_{(A)} \, \frac{\partial h_{ij}^{(1)}}{\partial \mathbf{R}} +
    \frac{1}{2} \sum_{ijkl} \Gamma^{ijkl}_{(A)} \, \frac{\partial h_{ijkl}^{(2)}}{\partial \mathbf{R}} .
    \label{eq:electronic_force}
\end{equation}
This requires the derivative of the 1- and 2-electron Hamiltonian matrix elements in the SAO basis. The AO-basis integral derivatives are widely available and we use the \emph{Libcint} library\cite{libcint2015} within the \emph{PySCF} package\cite{pyscf2018,pyscf2020}. It is then necessary to consider the nuclear derivative of the AO to SAO transformation matrix. This does {\em not} require the overhead of coupled-perturbed Hartree--Fock for the derivative of mean-field molecular orbital transformations, and can instead be simply constructed from first-order perturbation theory~\cite{bamieh2022-matrixperturbation}, described in more detail in the Supplementary Information of Rath et al.\cite{Rath24-evcont}.


\subsection{Nonadiabatic Coupling Vectors}
\label{sec:nac}

Transitions between different potential energy surfaces in mixed classical-quantum NAMD is governed by the nonadiabatic/derivative coupling vectors (NACs) between states. These describe the strength of the coupling and therefore the probability of transitioning. They are crucial for obtaining accurate internal conversion in NAMD (with intersystem crossing processes able to be described from the spin-orbit coupling in a similar framework), yet are often approximated due to the scarcity of analytic NACs in many electronic structure methods. The continuation scheme of this work allows for the extraction of these important quantities at all inferred geometries with little overhead.

The first order nonadiabatic coupling vectors between inferred states $A$ and $B$ are given, in its usual form, as
\begin{equation}
    \mathbf{d}_{AB}(\mathbf{R}) = \braket{\Psi_A(\mathbf{R})| \frac{\partial }{\partial \mathbf{R}} \Psi_B(\mathbf{R})}.
    \label{eq:nac}
\end{equation}
The derivative of both the subspace expansion coefficients, $x_b^{(A)}$ and the orbital basis need to be accounted for with two separate terms,
\begin{equation}
    \mathbf{d}_{AB} = \mathbf{d}_{AB}^\mathrm{coef} + \mathbf{d}_{AB}^\mathrm{orb}.
    \label{eq:nac_evcont}
\end{equation}
The coefficient dependent term can be found using the subspace expansion in Eq. \ref{eq:evcont_expansion}, the orthonormality condition between the inferred states, $\mathbf{x}_{(A)}^{T} \mathbf{Q} \, \mathbf{x}_{(B)} = \delta_{A,B}$ (where $\mathbf{Q}=Q_{ab}=\langle C_a | C_b \rangle$ is the overlap matrix between the training states), and the generalized Hellman-Feynman theorem, as
\begin{align}
    \mathbf{d}_{AB}^\mathrm{coef} &= 
    \sum_{ab} x_a^{(A)} \langle C_a | C_b \rangle \frac{\partial x_b^{(B)}}{\partial \mathbf{R}} = 
    \mathbf{x}_{(A)}^\dagger \mathbf{Q} \, \frac{\partial \mathbf{x}_{(B)}}{\partial \mathbf{R}} \\
    &= \frac{1}{E_B -E_A} \Big[ \mathbf{x}_{(A)}^\dagger \frac{\partial \mathcal{H}}{\partial \mathbf{R}} \mathbf{x}_{(B)} \Big] 
    \\
    &= \frac{1}{E_B -E_A} \Big[ \sum_{ij} \gamma^{ij}_{(AB)} \, \frac{\partial h_{ij}^{(1)}}{\partial \mathbf{R}} +
    \frac{1}{2} \sum_{ijkl} \Gamma^{ijkl}_{(AB)} \, \frac{\partial h_{ijkl}^{(2)}}{\partial \mathbf{R}} \Big] 
\end{align}
where the transition-RDMs between two different inferred states $A$ and $B$ are computed as:
\begin{equation}
    \gamma^{ij}_{(AB)} = \sum_{a,b}^M x_a^{(A)^*} \, \gamma^{ij}_{ab} \, x_b^{(B)},
\end{equation}
and
\begin{equation}
    \Gamma^{ijkl}_{(AB)} = \sum_{a,b}^M x_a^{(A)^*} \, \gamma^{ijkl}_{ab} \, x_b^{(B)}.
\end{equation}

Following a similar procedure to analytic CASSCF nonadiabatic coupling vectors\cite{Lengsfield84-NAC_CASSCF,Fdez16-NAC_CASSCF}, the orbital contribution, $ \mathbf{d}_{AB}^\mathrm{orb}$ is formulated as:
\begin{align}
    \mathbf{d}_{AB}^\mathrm{orb} &=  \sum_{ij} \gamma^{ij}_{(AB)}  \braket{ \chi_i | \frac{\partial \chi_j}{\partial \mathbf{R}} },
    \\
    &= \sum_{ij} \gamma^{ij}_{(AB)} \Big[ Z_{\alpha i} S_{\alpha\beta} \frac{\partial Z_{\beta j}}{\partial \mathbf{R}} + Z_{\alpha i} \braket{ \phi_{\alpha} | \frac{\partial \phi_\beta}{\partial \mathbf{R}} } Z_{\beta j} \Big]
\end{align}

\noindent where $Z_{\alpha i}$ is the rotation matrix from AO, $\phi_{\alpha}$ to SAO, $\chi_i$ representation.
These quantities are already computed during the evaluation of forces with the main difference being the contraction with t-RDMs, $\gamma^{ij}_{(AB)}$ instead of the RDMs, $\gamma^{ij}_{(A)}$, of the inferred states. The only additional quantity that is needed are the AO derivative coupling integrals, $\braket{ \phi_{\alpha} | \frac{\partial \phi_\beta}{\partial \mathbf{R}} }$, for the orbital contribution to NACs, which are readily available within the \emph{Libcint} library~\cite{libcint2015}.


\subsection{Proof-of-principle}

\begin{figure*}
\centering
  \includegraphics[width=\textwidth]{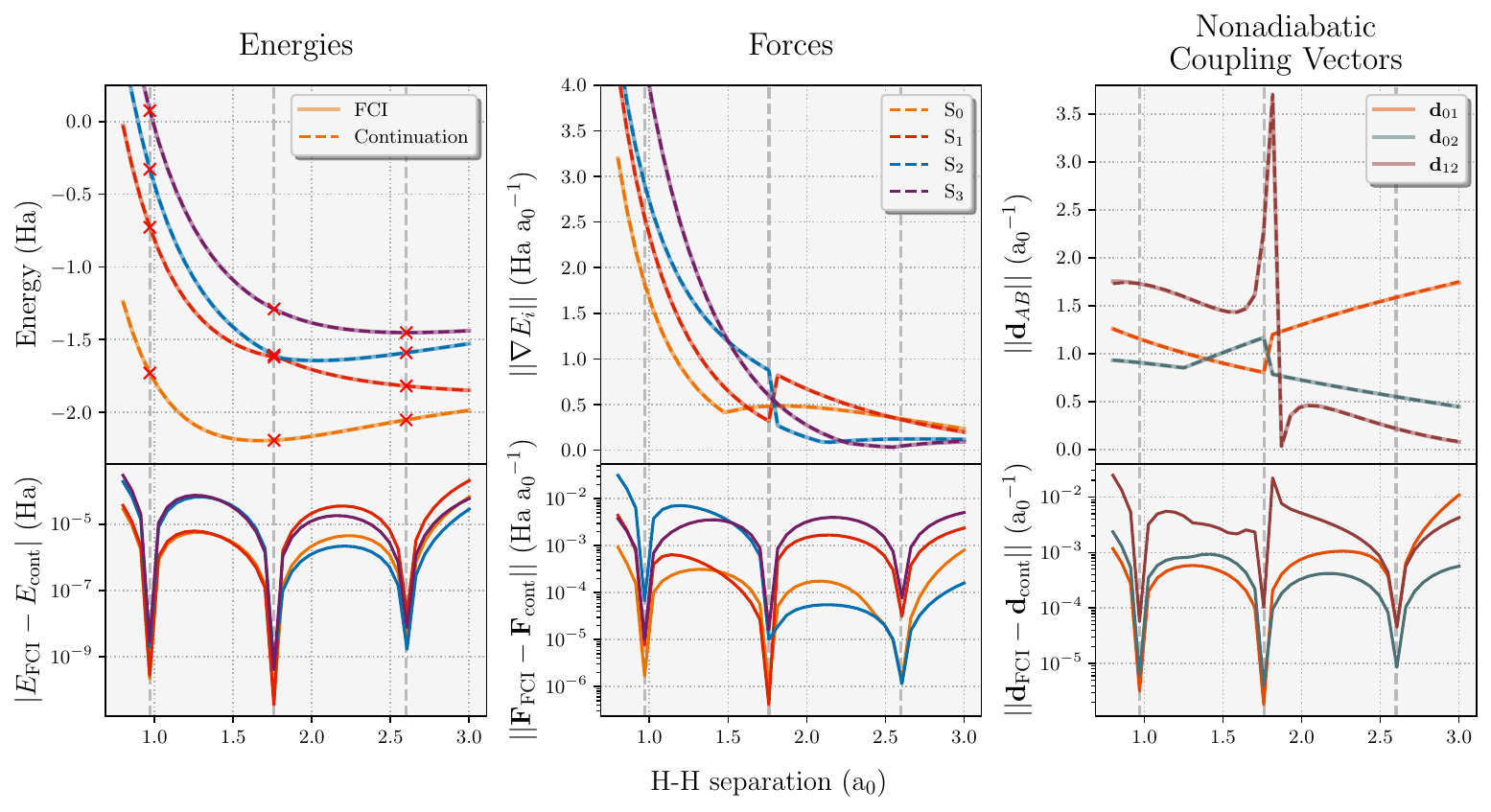}
  \caption{Smooth interpolation of ground and excited state energies, forces and nonadiabatic coupling vectors for the linear equidistant 4-atom hydrogen chain, with just three equally-spaced training FCI calculations (STO-3G basis). Red crosses represent the energies of the training wavefunctions used in eigenvector continuation (dashed lines), compared to FCI (solid lines). The absolute error of the eigenvector continuation for these properties is displayed in the bottom panels.}
  \label{fig:NAC}
\end{figure*}

The evaluation of interpolated multi-state energies, forces and NACs are particularly efficient in this scheme, given both the variationality of the inferred states as well as the required tRDMs between training states.
To validate the accuracy of this approach in a small system, we map out the multi-state potential energy surface of the four-atom linear hydrogen atom chain for the symmetric stretch in Fig.~\ref{fig:NAC}. In addition to energies, both the state-specific forces and NACs are computed, and their absolute error compared to FCI for all states. It should be noted that eigenvector continuation is exact with respect to FCI (to numerical precision) for all observables at the training geometries. At these points, the approach will simply select the interpolated wavefunction to directly be the FCI training point at that geometry, which minimizes the energy by definition. In addition, the energies computed from eigenvector continuation can never be lower than the FCI energies due to its variational nature. The absolute errors shown in bottom panels of Fig.~\ref{fig:NAC} also highlight the smoothness of the properties in the proposed scheme as there are no pathological sharp spikes in their error. With just three FCI calculations chosen from equally-spaced training geometries, the whole potential energy surface is predicted to well below chemical accuracy. The fact that the $S_0$ force does not go to zero at the minimum is just a reflection of the fact that only the symmetric stretching coordinate is considered. Furthermore, this coordinate exhibits a state crossing between $S_1$ and $S_2$, which is indicated by the appropriate divergence of the NAC between these states ($\mathbf{d}_{12}$).

\section{Nonadiabatic Molecular Dynamics on Interpolated Potential Energy Surfaces}

The motivating example of to justify the development of this multi-state interpolation scheme was for NAMD (though certainly not the only application that can be envisaged). In this section we consider the use of these interpolated energies, forces and NACs in order to propagate nuclear coordinates through time from initially excited electronic states. In this section, we rigorously benchmark against exact (FCI) dynamics, using this comparison to develop an active learning scheme for a selection of a compact training set to minimize the number of high-level calculations required across the trajectory. 



\subsection{Fewest-Switches Surface Hopping}
\label{sec:surfacehopping}

We briefly review FSSH\cite{Tully90-FSSH} as our NAMD method of choice, and one of the most widely used mixed quantum-classical dynamics approaches for studying photo-induced dynamics. The scheme allows electrons to relax through different electronic levels and transfer their energy to nuclear motion. The approach still relies on the adiabatic representation of potential energy surfaces, with the time-dependent electronic wavefunction described as a superposition over these low-energy states. The adiabatic states can be transformed into diabatic states via different diabatisation approaches if the propagation is done using the diabatic representation, but we will focus on the use of the adiabatic one, which is the most popular approach. The nuclear propagation is nevertheless performed according to a single electronic state, but this choice of state varies according to a stochastic hopping to introduce the nonadiabaticity in this propagation. Nuclear and electronic degrees of freedom therefore remain separated, with the nuclear dynamics propagated classically according to the force, $\mathbf{F}_A(\mathbf{R})$ due to potential generated by the adiabatic state $E_A$. 

The electronic dynamics are propagated by the time-dependent Schr{\"o}dinger equation as a superposition over the adiabatic surfaces at a given nuclear geometry as
\begin{equation}
    i\hslash \frac{d\varphi_A}{dt} - E_A \varphi_A + i\hslash \sum_{B} \mathbf{d}_{AB} \cdot \mathbf{v}  \, \varphi_B = 0,
    \label{eq:TDSE}
\end{equation}
where $\varphi_A$ are the expansion coefficients of the nonadiabatic electronic wavefunction in the basis of adiabatic states, $E_A$ is the energy of adiabatic state $A$, $\mathbf{d}_{AB}$ is the nonadiabatic coupling vector between the states as introduced in section \ref{sec:nac} and $\mathbf{v}$ is the current nuclear velocity. The coefficients $\varphi_A$ are updated at each step to account for decoherence corrections~\cite{Granucci07-decoherence}.
The population of the adiabatic state $A$ in the nonadiabatic wavefunction is given by $|\varphi_A(t)|^2$ at a given timestep. The probability of hopping between different adiabatic surfaces for the nuclear propagation is given by
\begin{equation}
    P_{A \rightarrow B} = \max \Big[0, -\frac{2\Delta t}{|\varphi_A|^2} \, \mathrm{Re}(\varphi_B \varphi_A^*) \, \mathbf{d}_{AB} \cdot \mathbf{v}   \Big]
\end{equation}
where $\Delta t$ is the timestep used in the integration of the nuclear propagation.
This probability is realised stochastically, with the hopping between state $A$ and $B$ occurring according to the uniform random number $\eta \in [0,1)$ if
\begin{equation}
    \sum_{C=1}^{B-1} P_{A \rightarrow C} < \eta \leq \sum_{C=1}^{B} P_{A \rightarrow C}.
\end{equation}
The nuclear velocities are rescaled according to their NACs after hopping to conserve the total energy over the trajectory~\cite{Barbatti2021-velocityadjustment}.

In this work we used the \emph{Newton-X}\cite{NewtonX14,NewtonX22} package to perform the FSSH simulations, interfaced to our eigenvector continuation code which was called to obtain all interpolated electronic structure properties, and NACs between all pairs of interpolated states were included for the propagation of the electronic wave function and stochastic hopping. This also relied on \emph{PySCF} for the computation of the required Hamiltonian, derivative integrals and FCI training data where used~\cite{pyscf2018,pyscf2020}. This workflow was also initially benchmarked against the FCI Newton-X interface to OpenMolcas~\cite{openMOLCAS2019,Merritt2023-openMOLCAS_NX_interface}. A 5\textsuperscript{th} order multistep integrator of Butcher\cite{Butcher65-multistepODEintegrator} was used to integrate Eq. \ref{eq:TDSE} with 20 multisteps. Decoherence corrections were applied through the simplified decay of mixing approach\cite{Granucci07-decoherence} with a decay parameter of 0.1 Ha. All trajectories were checked for energy conservation.

\subsection{Benchmarking against exact trajectories}

\ifjcp
\begin{figure*}
\centering
\includegraphics[width=0.8\textwidth]{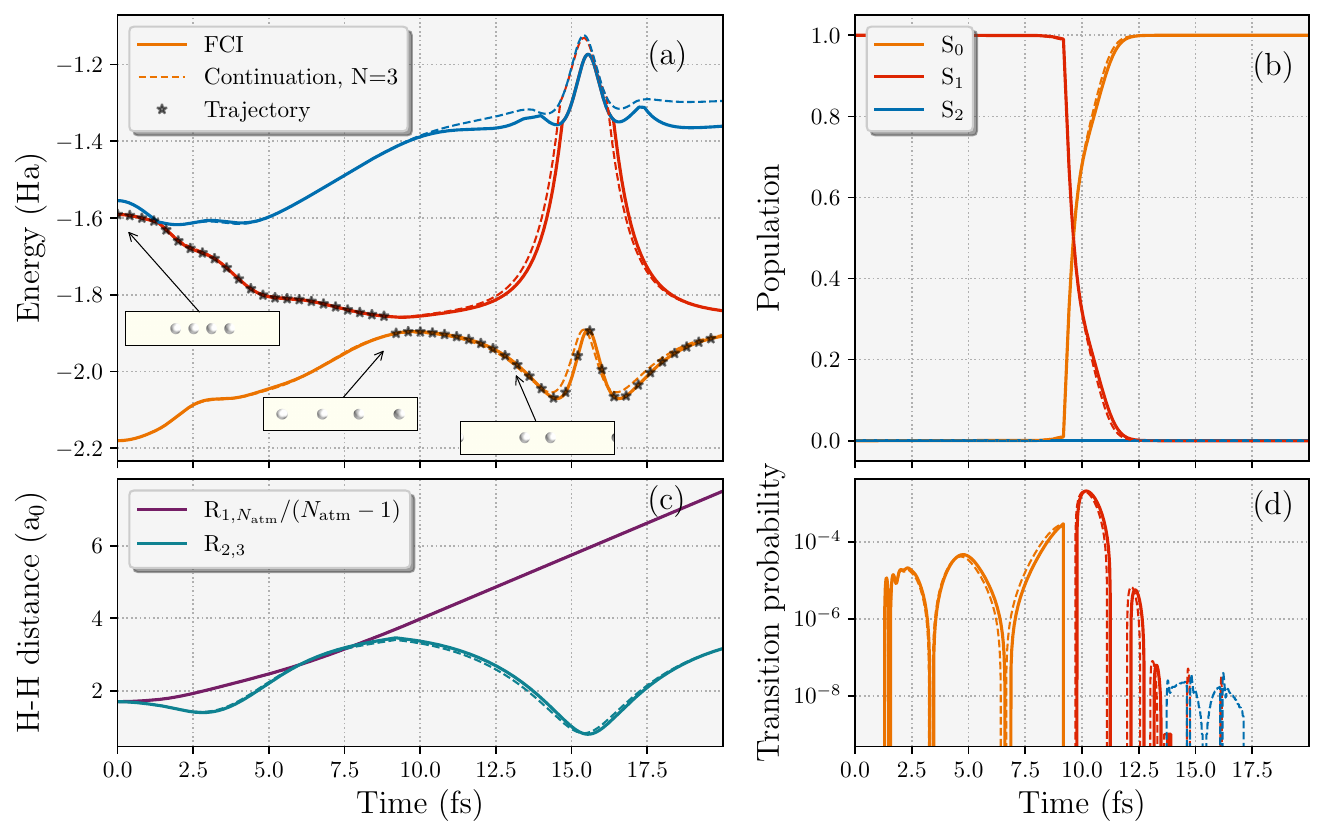}
  \caption{Comparison of the interpolated trajectory (dotted lines) against the exact results (solid lines) of a four-atom hydrogen chain starting in the equilibrium equidistant $S_1$ state of a STO-3G basis. (a) Adiabatic energies along the trajectory (shown with black stars) with a nonadiabatic hop to the ground state. Insets showing the nuclear geometries. (b) Populations of propagated electronic wavefunction over the low-energy adiabatic states considered. (c) The average interatomic distance of the full chain ($R_{1,N_{\mathrm{atm}}}/(N_{\mathrm{atm}}-1)$) and the distance between the middle hydrogens ($R_{2,3}$). (d) The transition probability from the current state of the trajectory to the other two states. The training wavefunctions used in the interpolation are taken from the three equidistant geometries shown in Fig.~\ref{fig:NAC}.}
  \label{fig:benchmark_NAMD}
\end{figure*}
\else
\begin{figure}[h]
\centering
\includegraphics[width=0.8\textwidth]{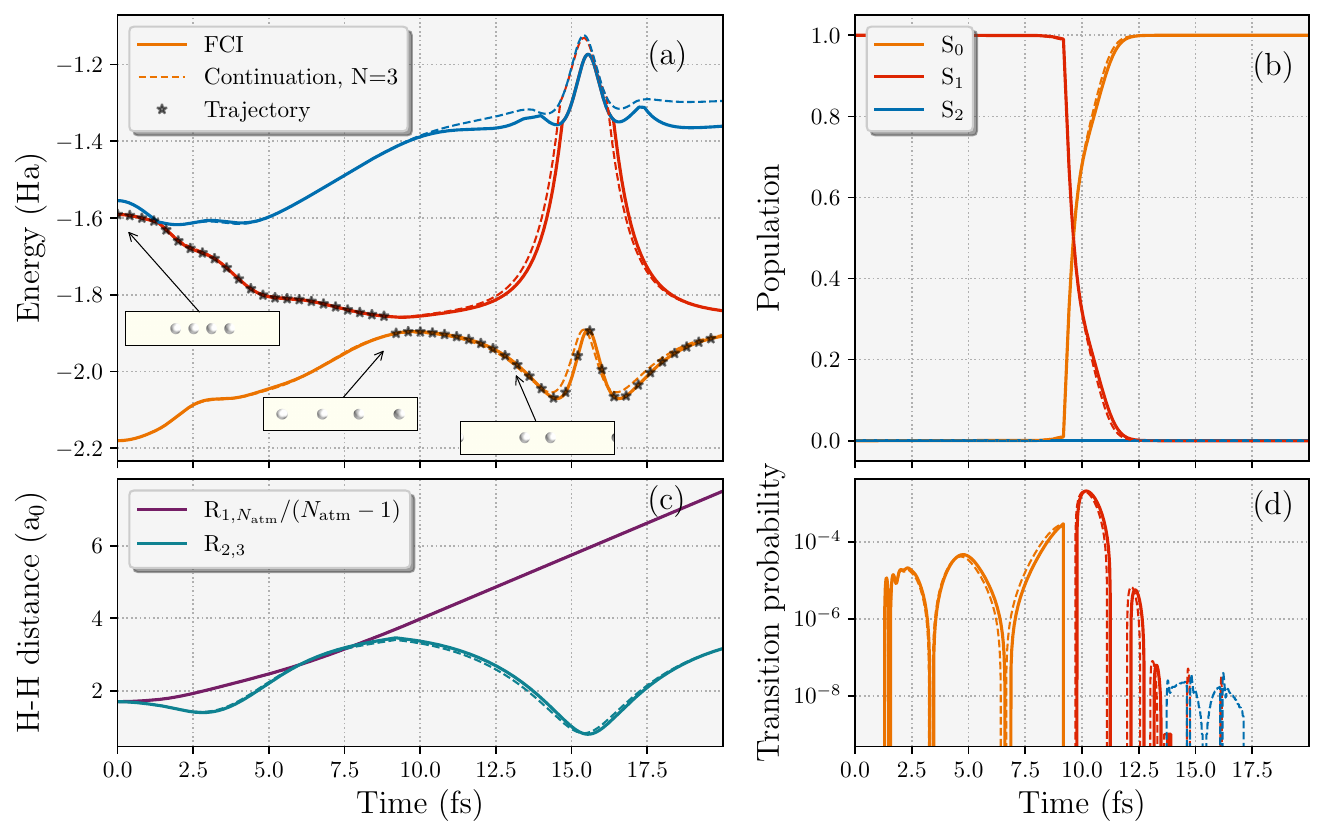}
  \caption{Comparison of the interpolated trajectory (dotted lines) against the exact results (solid lines) of a four-atom hydrogen chain starting in the equilibrium equidistant $S_1$ state of a STO-3G basis. (a) Adiabatic energies along the trajectory (shown with black stars) with a nonadiabatic hop to the ground state. Insets showing the nuclear geometries. (b) Populations of propagated electronic wavefunction over the low-energy adiabatic states considered. (c) The average interatomic distance of the full chain ($R_{1,N_{\mathrm{atm}}}/(N_{\mathrm{atm}}-1)$) and the distance between the middle hydrogens ($R_{2,3}$). (d) The transition probability from the current state of the trajectory to the other two states. The training wavefunctions used in the interpolation are taken from the three equidistant geometries shown in Fig.~\ref{fig:NAC}.}
  \label{fig:benchmark_NAMD}
\end{figure}
\fi

We first consider a NAMD trajectory of four Hydrogen atoms in a linear configuration, released from an initial equilibrium equidistant nuclear configuration with zero velocity of all nuclei and in the first electronically excited state (Fig~\ref{fig:benchmark_NAMD}). Three electronic states (S$_0$, S$_1$ and S$_2$) were interpolated in the simulations. The electronic wavefunction started in S$_1$ and the trajectory was propagated with a timestep for the nuclear dynamics of 0.05 fs in a minimal STO-3G basis. The $S_1$ state is fully dissociative, ensuring that all atoms initially move apart from each other (albeit at different speeds, and with the middle atoms initially moving closer). At $\sim8.5$ fs, it can be seen in Fig.~\ref{fig:benchmark_NAMD} that a jump to the ground state occurs, the nuclear velocities are rescaled, and the ground state potential surface is one that promotes dimerization.
However, the velocity of the end hydrogen atoms is already too high for the attractive potential to overcome and they continue to dissociate, increasing the length of the chain linearly with time. In contrast, the middle two hydrogens are able to come back together an dimerize, as seen in Fig. \ref{fig:benchmark_NAMD}(c). The end product is a vibrating H$_2$ molecule (with period $\sim 13$ fs) in the middle, with two individual atoms fully dissociating from the chain, rather than the perhaps anticipated scenario of two hydrogen dimers as the end result.

We now consider the performance of the eigenvector continuation compared to the exact propagation in this toy system, where we select FCI training states only at the same \emph{three equidistant} and evenly-spaced geometries as shown in Fig.~\ref{fig:NAC}. The key question is whether these non-equidistant, semi-dissociated nuclear geometries visited in the trajectory are well described in the scheme compared to FCI with these somewhat unrepresentative training points.
On top of that, we want to see if it can capture the correct internal conversion time to the ground state as small changes in the stochastic hopping transition can result in significant differences in the dynamics. Fig.~\ref{fig:benchmark_NAMD} demonstrate the success of the continuation even without any particular care in the training geometry selection. Despite the simplicity of this four-electron system, nine FCI training wavefunctions computed at three equidistant geometries seem to represent the low-energy Hilbert space (with a total of 256 Slater determinants) at \emph{all} the relevant geometries. In addition to the energies, populations and nuclear geometries, the transition probabilities between different adiabatic states in Fig.~\ref{fig:benchmark_NAMD} (d) are also captured to a very high quantitative accuracy. This indicates the methods capabilities to extract accurate statistical averages over multiple surface hopping trajectories. It could perhaps be argued that the success here is only reflective of the simple ratio of the Hilbert space size to number of training states. We therefore address this question for larger systems in Sec.~\ref{sec:largesys}.






\subsection{Active learning of training geometries}
\label{sec:active-learning}

Before extending to more complex systems, it is worth revisiting the determination of the optimal geometries from which electronic wave functions should be included as training states in the interpolation. This is an important step to maintain a compact representation of the relevant low-energy electronic space through MD trajectories and extend the prospects of the scheme to more realistic and complex systems over larger nuclear phase spaces. 
To achieve this, we iteratively add states from different training geometries to achieve a `black-box' procedure which can systematically converge the interpolated electronic surfaces of the relevant phase space in a self-consistent fashion, without relying on external guidance or information. This is done by starting from a single (or small number of) initial geometries in the training set, running an MD trajectory interpolating from this data and selecting the next geometry to add based on some heuristic estimation of the changing error for the geometries visited over the run. This process can be repeated until a desired level of convergence. This iterative process is usually referred to as `active' or on-the-fly learning as new geometries are chosen based on the information obtained from existing geometries.

The crucial component of any active learning strategy is the selection criteria for the new data for the training set at each iteration. It is common to use a distance metric to choose the data that is most different from the ones in the training set. Since the main inputs for our inference are the 1- and 2-electron integrals at the geometries of interest, which determine all electronic states, we employed this `Hamiltonian distance':
\ifjcp
\begin{align}
    D_\mathrm{min} (\mathbf{R}) &= \min_{\mathbf{R}_t \in \mathrm{train}} \Bigg[ \sum_{ij} |h^{(1)}_{ij}(\mathbf{R}) - h^{(1)}_{ij}(\mathbf{R}_t)|^2 \nonumber 
    \\ &+ \frac{1}{2} \sum_{ijkl} |h^{(2)}_{ijkl}(\mathbf{R}) - h^{(2)}_{ijkl}(\mathbf{R}_t)|^2 \Bigg], \label{eq:hamdist}
\end{align}
\else
\begin{equation}
    D_\mathrm{min} (\mathbf{R}) = \min_{\mathbf{R}_t \in \mathrm{train}} \Bigg[ \sum_{ij} |h^{(1)}_{ij}(\mathbf{R}) - h^{(1)}_{ij}(\mathbf{R}_t)|^2 + \frac{1}{2} \sum_{ijkl} |h^{(2)}_{ijkl}(\mathbf{R}) - h^{(2)}_{ijkl}(\mathbf{R}_t)|^2 \Bigg], \label{eq:hamdist}
\end{equation}
\fi
as our distance metric where $\mathbf{R}_t$ are the geometries in the existing training set and, $h^{(1)}_{ij}$ and $h^{(2)}_{ijkl}$ are the electron integrals. This can be simply evaluated along with the inference over the MD trajectory, and satisfies the important property that the metric is zero for the interpolation at existing training geometries.

This metric was applied in a previous publication to converge the training states for ground state dynamics with eigenvector continuation, where the geometry with largest $D_\mathrm{min}$ over the trajectory was solved with the electronic structure solver and added to the training data for the next MD run, iteratively improving the trajectory until convergence~\cite{Rath24-evcont}. This ensures that geometries from the trajectory are added to the training set that are `furthest' (in this Hamiltonian distance sense) from the nearest training state. However, simply taking the maximum $D_\mathrm{min}$ over the entire trajectory doesn't yield the fastest convergence, as these points tend to correspond to the final geometries visited in the trajectory, where the furthest parts of the phase space are being explored. A better measure is to consider the turning points, i.e. peaks in the Hamiltonian distance, over the trajectory, as they signal two important scenarios. Firstly, they can point to true dynamical extrema such as the minimum and maximum separation of a nuclear oscillations. For an optimal interpolation (rather than extrapolation) of that motion, inclusion of these extrema is beneficial. The second scenario is when the continuation trajectory is in an explorative phase (i.e. we are searching for additional training geometries required, rather than running the final converged trajectory). As the scheme introduced in this paper necessarily overestimates the energies of atomic geometries in regions of phase space far from the training data due its variational nature, the MD has an inductive bias away from these spuriously high-energy regions. Following this, the trajectory hits an overestimated energy barrier as it moves towards these poorly described geometries and bounces back to a known part of the phase space. Thus, peaks in Hamiltonian distance also hint at these regions of the nuclear phase space where the addition of training data would be particularly beneficial to optimally improve the overall accuracy of the interpolation.

As the nuclear phase space explored during the dynamics changes as more training data is added to the model (since the inferred electronic potential changes), consideration needs to be made as to the trade-off between exploration and exploitation in the learning.
Therefore some bias needs to be included to the heuristic metric for the choice of the training geometry to include, to preferentially select geometries from earlier times in the trajectory, before it has diverged too much from the exact converged path. This avoids unnecessary additional electronic structure calculations to include training geometries which may be unrepresentative of the phase space explored in the final converged trajectory.
To account for this, the peaks in $D_\mathrm{min}$ are weighted according to equation~\ref{eq:weighted_hamdist}, to bias the selection of training geometries that occur earlier in the trajectory,
\begin{equation}
    \mathbf{R}_\mathrm{add} \, = \, \argmax_{\mathbf{R}_i \, \in \, \mathrm{peaks}} \; \Bigg( \frac{D_\mathrm{min} (\mathbf{R}_i) }{[ \, \nicefrac{t_i} {t_\mathrm{sim}} \,]^x} \Bigg)
    \label{eq:weighted_hamdist}
\end{equation}
where $\mathbf{R}_i$ is a nuclear geometry that corresponds to a peak in $D_\mathrm{min}$, $t_i$ is the time that geometry occurs in the trajectory and $t_\mathrm{sim}$ is the total length of the trajectory.

Here, the exponent $x$ is a hyperparameter that determines the degree of weighting towards earlier geometries. At its limiting cases, it can reduce to either selecting the highest peak ($x=0$) or the first peak ($x \rightarrow \infty$) in the Hamiltonian distance, $D_\mathrm{min}$. We found $x=3$ to result in a reasonably fast convergence, for the hydrogen chains studied in this work, providing a good balance between exploring the most unfamiliar geometries and including earlier peaks to ensure systematic convergence of the true trajectory from earlier to later times.

\begin{figure*}
\centering
    \includegraphics[width=\textwidth]{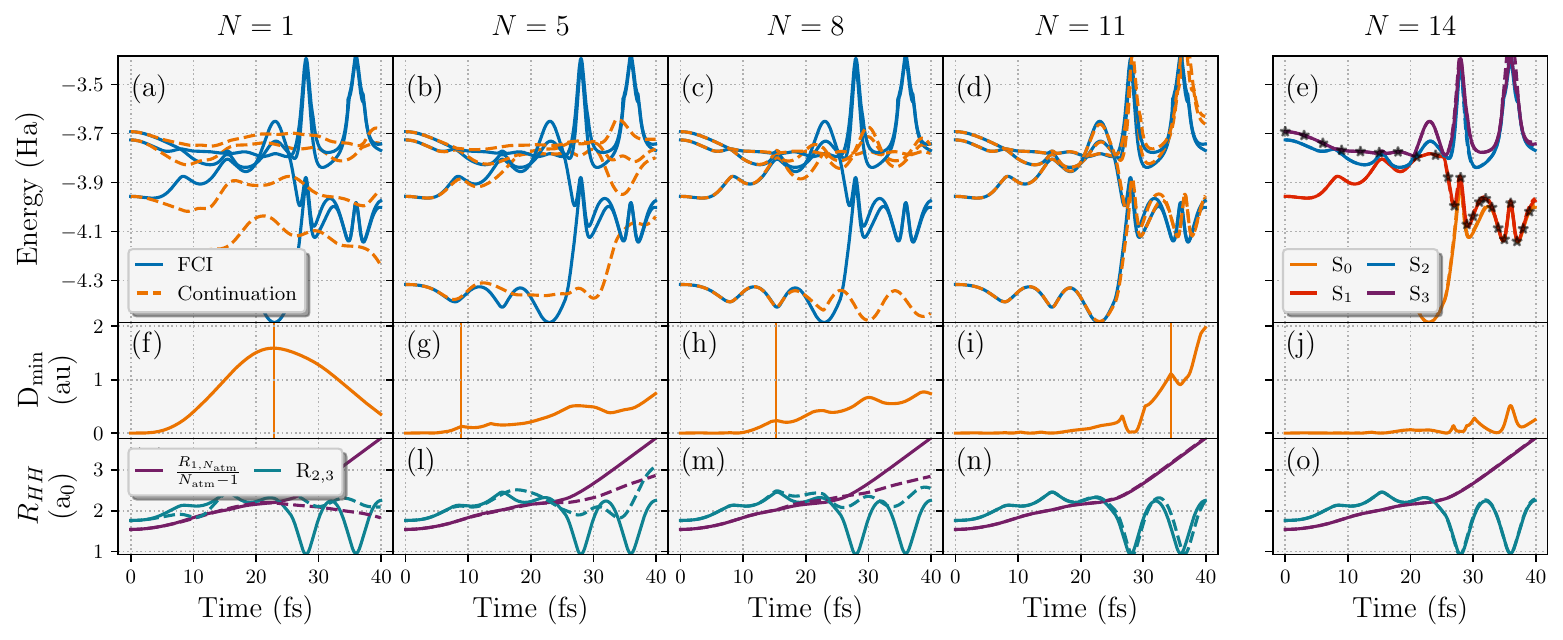}  
  \caption{Convergence of the active learning scheme for the nonadiabatic molecular dynamics trajectory of an eight-atom hydrogen chain with respect to number of geometries selected for training the model ($N$). Top panels (a-e) show the adiabatic energies along this trajectory (exact as blue solid lines, interpolated from $N$ training FCI calculations as orange dashed lines). Bottom panels (k-o) display the scaled end-to-end distance of the chain, $R_{1,N_\mathrm{atm}}/(N_\mathrm{atm}-1)$ and the distance between the 2\textsuperscript{nd} and 3\textsuperscript{th} hydrogen atoms along the chain, $R_{2,3}$. Middle panels (f-j) show the minimum Hamiltonian distance (Eq.~\ref{eq:hamdist}) over the geometries along the continuation trajectory with respect to the training geometries available at that iteration. The vertical lines in this panel represent the next geometry along the trajectory selected for inclusion in the model, based on the metric discussed in Eq.~\ref{eq:weighted_hamdist}.}
  \label{fig:active-learning}
\end{figure*}

This learning strategy is demonstrated in Fig. \ref{fig:active-learning} for converging the NAMD trajectory over five inferred states with respect to number of geometries used in the training ($N$) of an eight-atom hydrogen chain, using the five lowest-energy FCI states from each training geometry. The interpolated surfaces with the shown number of training geometries ($N$) are shown with orange dashed lines in panels (a)-(e), reflecting both the error in the interpolation and the divergence of the trajectories as a result, compared to the exact adiabatic surfaces from FCI (solid blue lines). All adiabatic surfaces were restricted to be from the same spatial symmetry as the ground state. The trajectory was started from the third excited state, $S_3$, at the equilibrium equidistant linear chain geometry with zero nuclear kinetic energy and propagated with a timestep of 0.1 fs in the minimal STO-3G basis. The interpolated trajectory agrees almost exactly with the FCI trajectory by the time $N=14$, as shown in \ref{fig:active-learning}(e). In this, the propagation follows $S_3$ for around 20 fs which pushes the system to form two separate H$_4$ clusters, before the trajectory hops to the $S_1$ state pushing the middle hydrogens of each of these H$_4$ clusters to form energetic dimers while ejecting the end hydrogens in opposite directions. Both of these can be seen in \ref{fig:active-learning}(o) where the dimerization of the second and third hydrogen atoms and the rapid increase in the end-to-end distance is observed just after 20 fs. After this, the system stays in $S_1$ for 15 fs before jumping to the ground state around 35 fs. This, in turn, stabilises the dimers and allow the formation of another dimer in the middle between the fourth and fifth hydrogens that were ejected from their respective clusters. This leads to a final configuration of three vibrating H$_2$ dimers in the middle with two atomic hydrogen atoms dissociating from the chain. This complex NAMD trajectory explores a lot of different regions of the nuclear phase space that would be difficult to select as training states \emph{a priori}.

We can also use this to analyze the convergence of the active learning scheme for selected numbers of geometries up to the converged $N=14$ trajectory, with the Hamiltonian-distance metric of Eq.~\ref{eq:hamdist} shown in Fig.~\ref{fig:active-learning} panels(f)-(j). The geometries selected for subsequent inclusion in the training set from the path of the inferred trajectory are indicated by vertical lines in these panels, according to Eq.~\ref{eq:weighted_hamdist}. It can be seen that the peaks in the Hamiltonian distance serve as qualitative indicators for points of divergence between the trajectory on the inferred potential and the FCI trajectory. This is especially evident for $N=5$ and $N=8$ trajectories where the data selection in Fig.~\ref{fig:active-learning}(g) and Fig.~\ref{fig:active-learning}(h) corresponds to the geometries where the divergence starts in Fig.~\ref{fig:active-learning}(b) and Fig.~\ref{fig:active-learning}(c), respectively. Notably, the $N=5$ trajectory underscores the significance of weighting the selection to earlier times since selecting the maximum $D_\mathrm{min}$ would have lead to configurations unexplored within the true trajectory. This allows for a systematic convergence of data selection towards the true trajectory from earlier to later times. Moreover, prioritizing geometries in this way also facilitates the exploration of relevant regions in the phase space at later times in the MD, as illustrated for $N=11$ in Fig.~\ref{fig:active-learning}(d). 
The focus on improving the accuracy of the model initially for earlier times ensures that the transition region between $S_1 \rightarrow S_0$ can be predicted correctly before trying to iteratively improve the later time sampling of the relevant post-transition ground state dynamics. These are targeted from $N=11$ onwards (see the data selection at 35 fs in Fig.~\ref{fig:active-learning}(i)) when relevant parts of the phase space in these trajectories are being sampled. 
Overall, the proposed heuristics appear to address the challenges of selecting a compact set of training geometries in a black-box and rapidly convergent fashion, minimizing the number of explicit electronic structure calculations required. This balances the challenges of considering both the time and the magnitude of the peak in the heuristic error estimate in the interpolated energy surfaces over the trajectories sampled. Overall, in this system qualitative convergence is achieved by the 11\textsuperscript{th} training geometry, with quantitative accuracy attained after a total of 14 training geometries. Considering both the nuclear phase space explored and electronic complexity of this trajectory, the proposed learning scheme can train and replicate the true trajectory with remarkably few FCI training points. We anticipate however that further improvements could be made by including a description of the inferred states themselves in these heuristics since only the Hamiltonian distance to a single training point is considered. This will be investigated in future work.



\section{Beyond Exact Methods: Hydrogen Chain Dynamics with DMRG Continuation} \label{sec:largesys}

\ifjcp
\begin{figure*}
\centering
  \includegraphics[width=\textwidth]{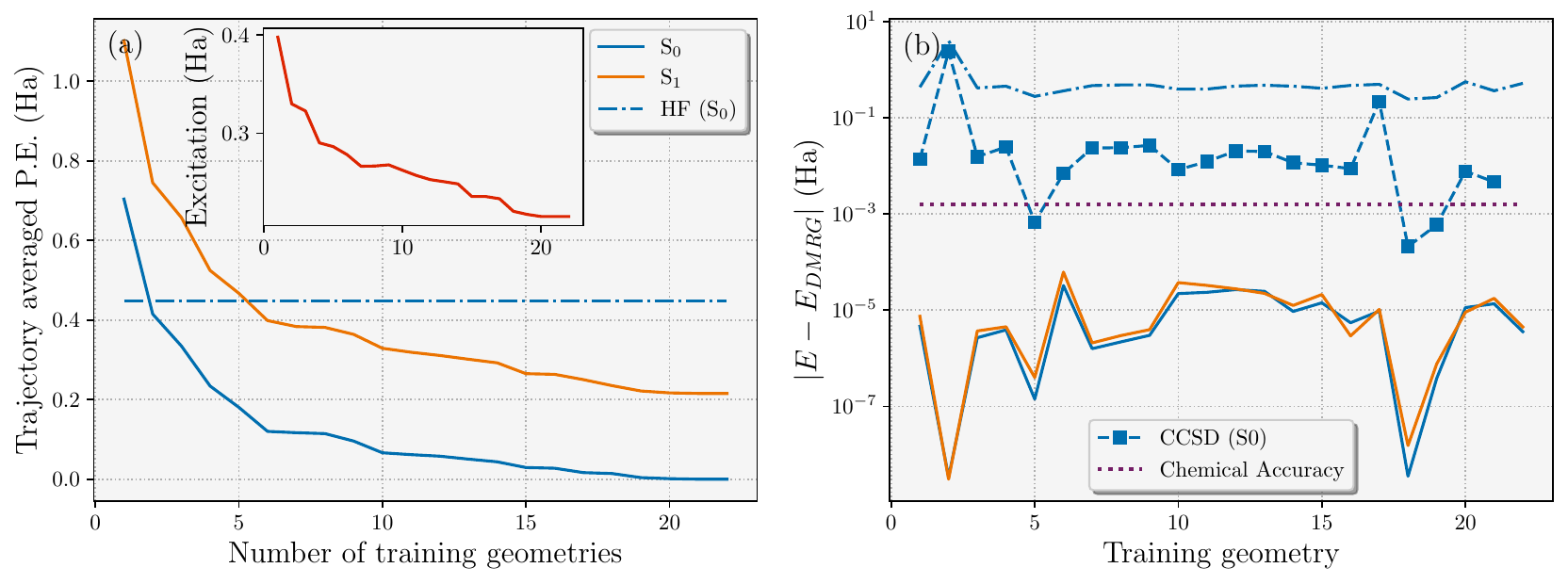}
  \caption{(a) Average variational energy of the $S_0$ and $S_1$ interpolated states over all geometries accessed in the final converged NAMD trajectory of the 28-atom hydrogen chain, as a function of number of DMRG training geometries used for the interpolated state. We can see smooth and systematic convergence of the DMRG-interpolated potential to mE$_h$ accuracy, with Hartree--Fock theory accuracy shown for comparison. The inset shows the convergence of the $S_0 \rightarrow S_1$ excitation energy. (b) Absolute energy error computed at the training geometries compared to the explicit DMRG used in the training. Dashed-dotted and dashed lines correspond to Hartree-Fock and CCSD comparison values for the ground state energies at these geometries. The error compared to the interpolated energy at the training geometries is close to numerically zero.}
  \label{fig:H28-convergence}
\end{figure*}
\fi

With the experience of previous sections benchmarking on exactly solvable hydrogen chains, we move on to larger lengthscales and timescales, where we can no longer compare to exact FCI results over timescales which would require a prohibitive number of calculations. Furthermore, accessing larger lengthscales require the use of approximate solvers for the training states. Nevertheless, we show that we can still have confidence and provable quantitative convergence to near-exactness in both the inferred electronic states and nonadiabatic trajectory based on the developed heuristics. 
Matrix product state wavefunctions optimized through \emph{ab initio} density matrix renormalization group (DMRG) methodology present itself as an highly accurate, systematically improvable approach to obtain \emph{near-exact} potential energy surfaces for system sizes beyond the reach of FCI~\cite{10.1063/1.4905329,10.1063/5.0050902}. By continuing our investigation of the nonadiabatic dynamics of longer hydrogen chains, we can explore larger nuclear phase spaces, while the quasi one-dimensional topology of the strongly correlated electronic structure is particularly efficient for DMRG solvers. These linear hydrogen systems have recently come to the fore as paradigmatic benchmark systems for a wide range of electronic structure methods, as a step towards extended condensed phase systems exhibiting a surprisingly rich phase diagram.\cite{Motta17-hydrogenchain,Motta20-hydrogengroundstate} We argue that their nonadiabatic dynamics would be unable to be explored in any other way, with other electronic structure methods unsuitable due to the lack of an obvious CAS, or unable to access the required time or lengthscales in this study. 

In recent years, there has been progress in obtaining approximate DMRG-SCF excited state gradients and NACs\cite{Freitag2019-approximateNAC-DMRGCASSCF,Reiher20-DMRGexcited} as well as quantum dynamics simulations of realistic vibronic Hamiltonians with time-dependent DMRG\cite{Yao2018-tDMRG,Baiardi2019-DMRGdynamics,Ren22-TDDMRG}. However, running full NAMD simulations with \emph{ab initio} DMRG has practically been out of reach due to difficulties with exact analytical NACs, high computational cost and difficulty in ensuring a fully black-box and robust workflow over the many calculations. Here, we present the first NAMD simulation with fully \emph{ab initio} MPS wavefunctions accelerated through the eigenvector continuation. We obtain DMRG training data via the spin-adapted and multi-state implementation in the \emph{block2} library, which can also straightforwardly obtain the required tRDMs and overlaps between the MPS training states~\cite{Block2,10.1063/5.0050902,Dorando07-multistateDMRG}. 

We consider the 28-atom one-dimensional hydrogen chain in a STO-6G basis with FSSH, where we initialize the chain with an equilibrium equidistant nuclear configuration with zero nuclear kinetic energy, photo-excited to the first excited electronic $S_1$ state, and we consider the electron and nuclear dynamics over these two interpolated states. The training MPS wavefunctions were optimized to near-exact accuracy by exponentially increasing the bond dimension at each training geometry while decreasing the noise during the DMRG sweeps, simultaneously optimizing the ground and $S_1$ state. 
A timestep of 0.5 fs was used for the FSSH trajectories.

\ifjcp
\else
\begin{figure*}
\centering
  \includegraphics[width=\textwidth]{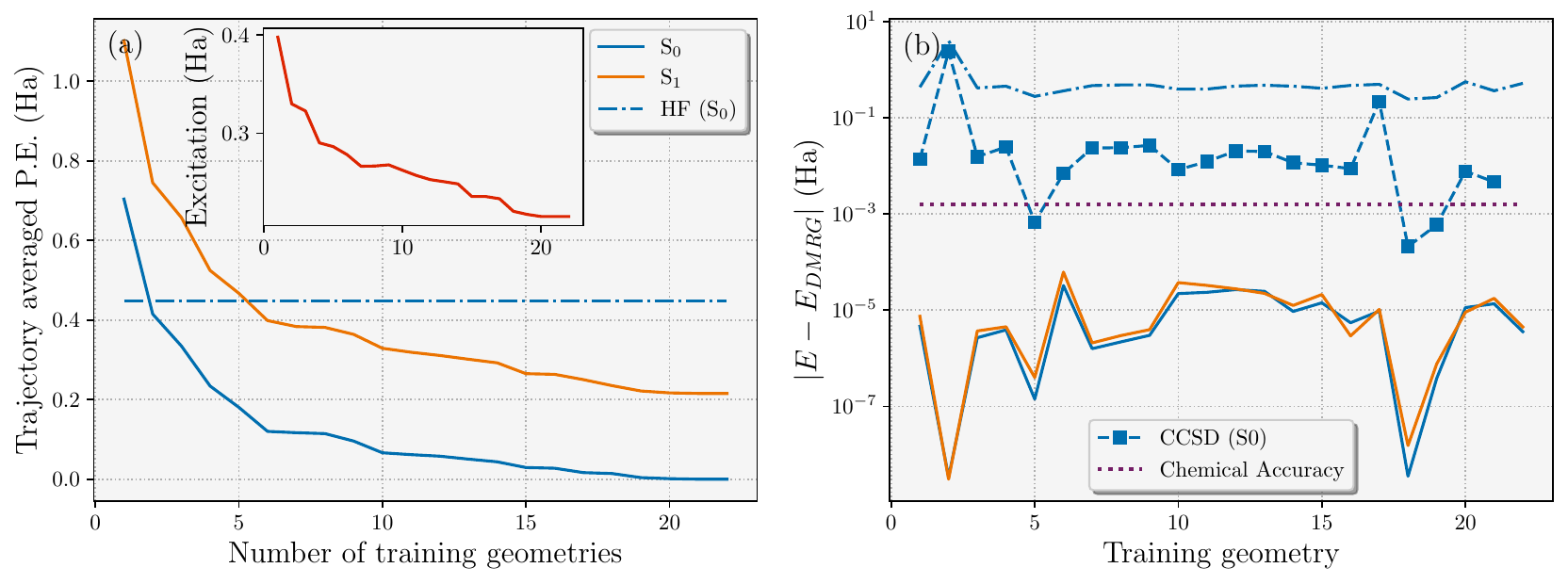}
  \caption{(a) Average variational energy of the $S_0$ and $S_1$ interpolated states over all geometries accessed in the final converged NAMD trajectory of the 28-atom hydrogen chain, as a function of number of DMRG training geometries used for the interpolated state. We can see smooth and systematic convergence of the DMRG-interpolated potential to mE$_h$ accuracy, with Hartree--Fock theory accuracy shown for comparison. The inset shows the convergence of the $S_0 \rightarrow S_1$ excitation energy. (b) Absolute energy error computed at the training geometries compared to the explicit DMRG used in the training. Dashed-dotted and dashed lines correspond to Hartree-Fock and CCSD comparison values for the ground state energies at these geometries. The error compared to the interpolated energy at the training geometries is close to numerically zero.}
  \label{fig:H28-convergence}
\end{figure*}
\fi

The active learning scheme described in Section~\ref{sec:active-learning} was applied to converge a single nonadiabatic molecular dynamics trajectory of the H$_{28}$ system that explores the relevant phase space and samples the dynamics on both S$_0$ and S$_1$ potential energy surfaces. It should be stressed that the training points selected by the scheme will \emph{not} (in general) be featured in the final trajectory, since they are taken from different potential energy surfaces as the inferred model is iteratively improved. This suggests that while it converges the trajectory from a single seed, it should exhibit little bias towards this single trajectory in the final data selection. Similar accuracy should be found over a stochastic ensemble of trajectories, with the selection relatively insensitive to the precise details of the trajectory. The convergence was achieved with $N=22$ geometries as shown in Fig.~\ref{fig:H28-convergence}(a). In this we show the convergence of the average variational energy of the S$_0$ and S$_1$ states with respect to $N$ over all geometries of the same final converged trajectory. We find that the addition of the final two training points changes this metric for \emph{both} states by less than chemical accuracy (1~kcal/mol) across the entire trajectory, indicating confidence in convergence with respect to the training data. It is also worth noting that only two training geometries were sufficient to surpass the variational energy of Hartree-Fock over the trajectory, despite no mean-field information considered in the interpolated states. The all-important $S_0 \rightarrow S_1$ average excitation energy over the trajectory also converges faster than the absolute energies of the states, as shown in the inset.

For some context as to the accuracy of the interpolated surfaces, Fig.~\ref{fig:H28-convergence}(b) illustrates the error in the interpolated $S_0$ and $S_1$ energies at the final $N=22$ training geometries, compared to the explicit DMRG used in its training, along with additional $S_0$ energy comparison to Hartree-Fock and Coupled Cluster Singles and Doubles (CCSD) theory. We find that the interpolated energies are exact (to numerical precision) compared to the DMRG calculations, as expected. However, at two geometries, the continuation scheme yields a slightly lower variational energy than the DMRG training energies for the $S_1$ state, indicated that a small further optimization of this state with DMRG would be possible by increasing bond dimension.
The H$_{28}$ system is close to an ideal system for DMRG due to its one-dimensional nature, so suboptimal convergence behavior was rare in this case. However, this showcases that convergence difficulties in the training for a particular geometry is not as problematic for the performance of the continued model, as it would likely be during a traditional MD trajectory. This is because the model will `borrow' electronic character from other training states to ensure that the surfaces remain smooth and therefore variationally improve upon unoptimized training states even at the training geometries themselves. This behavior was more evident when applied to the ground state dynamics of the Zundel cation reaction in the previous work~\cite{Rath24-evcont}. 


This unreliability in convergence of multiple single-point calculations can even be evidenced in comparison CCSD energies at the training geometries -- generally quite a robust electronic structure method, shown in Fig.~\ref{fig:H28-convergence}(b). Although it is well known that CCSD will struggle to capture the multi-reference nature of the more dissociative geometries due to its intrinsic single-reference formulation (geometry 2), it can also converge to the wrong state (geometry 17) or fail to converge at all (geometry 22) with simple default simulation parameters in widely used software packages. Isolated convergence issues are common when running lots of single-point calculations with different electronic character as are necessary in MD with many-body methodologies, and the proposed scheme ensures the smoothness of the potential energy surface and thus the physicality of dynamics. Even without these points however, CCSD would not reach chemical accuracy (even for energy \emph{differences}) across the trajectory.

\begin{figure*}
\centering
  \includegraphics[width=\textwidth]{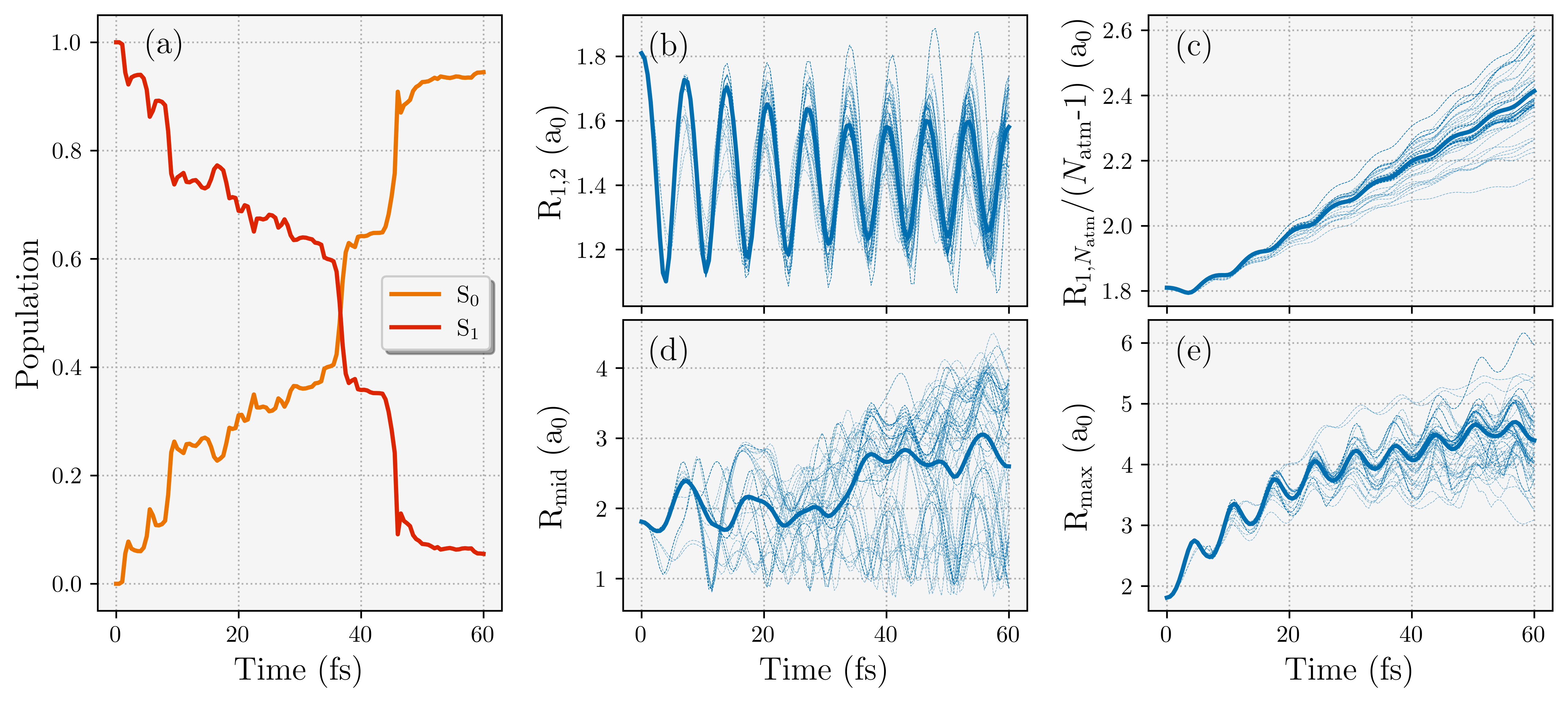}
  \caption{Nonadiabatic molecular dynamics simulations of a 28-atom hydrogen chain, starting from a stationary equally-spaced equilibrium configuration in the electronically excited $S_1$ state in a STO-6G basis. This required 12,000 energy, force and NAC calculations, inferred from 22 explicit DMRG training states. (a) Average electronic state population, (b) distance between the two end hydrogen atoms, (c) scaled end-to-end distance of the chain, (d) distance between the middle (14$^\mathrm{th}$ and 15$^\mathrm{th}$) hydrogen atoms, and (e) maximum separation between any two consecutive hydrogen atoms along the chain. The bold straight lines represent the average property over 100 trajectories while the thin dashed lines represent individual trajectories.}
  \label{fig:H28-multtraj}
\end{figure*}

Using these DMRG training wavefunctions for H$_{28}$ we find the dynamics over 60 fs for an ensemble of 100 trajectories with different random seeds, all starting at the equilibrium geometry with equally separated hydrogen atoms, in the $S_1$ electronic state. This is a total of 12,000 single-point calculations, all inferred from the same 22 DMRG training calculations. The averages over this ensemble of trajectories for the electronic population of each state and atomic distances are shown in Fig.~\ref{fig:H28-multtraj} to better understand the nature of the nonadiabatic molecular dynamics in this H$_{28}$ system.

The average electronic state populations are shown in Fig.~\ref{fig:H28-multtraj}(a), with the fastest rate of internal conversion between 30-50 fs after release, indicated in particular by two sharp jumps, with almost all trajectories in the ground $S_0$ state after this time.
In addition, there exists a handful of trajectories that hop back to the higher-energy $S_1$ state after hopping to $S_0$, especially among the trajectories that transition to $S_0$ early (before 10 fs), which can be seen by the dips in $S_0$ population around 5-15 fs.
Considering the nuclear geometries along the trajectories, it is clear that the hydrogen atoms at the ends of the chains rapidly dimerize for {\em all} trajectories, which is shown in Fig.~\ref{fig:H28-multtraj}(b). This contrasts with the H$_4$ and H$_8$ systems, where their terminal atoms were ejected before being able to dimerize. In the longer chain dynamics, the frequency and amplitude of these terminal dimer vibrations also seems to change with time, which differ from the ground state behaviour where a constant frequency and amplitude is maintained throughout the dynamics.
This might imply that the first excited state of H$_{28}$ is not as rapidly dissociative (or dissociative at all) unlike the smaller chains, or that the smaller excitation energy in this longer chain leads to faster decay to the ground state which favors dimerization. 


The end-to-end distance of the 28-atom hydrogen chain increases linearly with time as seen in Fig. \ref{fig:H28-multtraj}(c) with relatively little scatter in this rate over the different trajectories. The rate of this increase is very similar to the one observed in ground state hydrogen chain dynamics in Rath et al.~\cite{Rath24-evcont}. Note that the dynamics in that work started from a geometry that was 10\% stretched from the equilibrium, while this work starts from the first excited state of the equilibrium equidistant geometry. This suggests that the electronic energy gained by that initial stretch results in a equivalent chain length behaviour as a photo-excitation to S$_1$. 
Fig.~\ref{fig:H28-multtraj}(e) shows the maximum distance between any two neighbouring hydrogen atoms, which essentially describes the distance between the terminal dimers in the chain which move apart from each other at the fastest rate (i.e. the H$_2$-H$_3$ distance).


Unlike the consistent dimerization behavior at the ends of the chain, the behaviour in the middle of the chain seems to vary significantly depending on the trajectory, as seen in Fig.~\ref{fig:H28-multtraj}(d), exploring a large phase space. Continuous oscillations of some trajectories about 1.5 a$_0$ are visible, indicating that a portion of trajectories form dimers between the 14$^\mathrm{th}$ and 15$^\mathrm{th}$ hydrogen atoms. This would result in a frustrated dimerization of the overall chain, as full dimerization would require the 14$^\mathrm{th}$ and 15$^\mathrm{th}$ hydrogen atoms to dimerize with their other neighbours, not with each other. This therefore requires other non-dimer configurations in the chain for those trajectories, either forming trimers, unbound hydrogen atoms, or other larger hydrogen complexes.

We analyze the geometries that emerge in the last 5 fs of the 100 trajectories computed in this work, to identify different nuclear configurations that result. Of these trajectories, 27\% form perfect dimerization throughout the chain, where the distance within dimers is always smaller than the distance to their next neighbour, leaving the majority without this expected order. 39\% of all trajectories have hydrogen atoms being shared with two separate dimers where the bond separation within both of these dimers is less than the distance to their other neighbours. Only 9\% of trajectories can be characterized as having a `free' hydrogen where the distance to its neighbours is always more than the distance between the next consecutive hydrogens. Finally, the rest of the trajectories (25\%) don't fit any of these criteria and can be thought of as transition states between these configurations. Videos showcasing representative trajectories are available in the Electronic Supplementary Information.$^{\dag}$

The variation in the trajectories all appears simply due to the stochastic nature of the hopping to the ground state in the FSSH approach, which can result in significant deviations in this system and the exploration of different local minima in the nuclear phase space. A consideration of nonadiabatic models beyond FSSH would be interesting to see whether these differences persist, as well as further insights into the electronic behaviour over time (e.g. (transition) dipole moments and absorption spectra), which are all accessible from the interpolated states. In addition, there are a number of other physical mechanisms to consider, in particular the effect of proton tunnelling, as well as the incomplete nature of the basis set for the electronic states, which will be considered in future work.
While the electronic phase of the clamped equidistant hydrogen chain was found to exhibit surprisingly rich physics~\cite{Motta20-hydrogengroundstate}, it appears that the nuclear dynamics similarly exhibits significant intriguing and subtle physics. We suggest that this system could also be used as an effective benchmarking test bed for nuclear dynamics developments in correlated electron systems, as well as for the electronic structure solvers on which they so heavily depend.

\section{Conclusions and Outlook}

In this work, we have extended the powerful interpolation scheme for many-body wavefunctions to a multi-state approach, and demonstrated that this can be used to easily find analytic excited state gradients and nonadiabatic couplings between inferred states. This makes the approach ideal for accelerating nonadiabatic molecular dynamics calculations, and to straddle the gulf in accessible timescales which hinders the application of emerging electronic structure methods in this important domain. Crucially, the electronic structure interpolated through chemical space is necessarily smooth and systematically improvable, and we develop an active training protocol to iteratively ensure a compact and rapidly convergent number of training wavefunctions required for a subsequent MD. While the cost of each electronic structure training calculation is high, the subsequent inference is possible over significant timescales due to the non-iterative quartic scaling with system size in evaluating the properties of the inferred electronic model.

We demonstrate the approach with fewest-switches surface hopping on one-dimensional hydrogen chains, which exhibits a surprisingly rich dynamical behaviour and wide range of trajectories. For the largest H$_{28}$ system, we use $22$ training states as matrix product states from single-point DMRG calculations, and infer $12{,}000$ points on a smooth multi-state electronic surface converged to chemical accuracy from this DMRG training (including their analytic atomic forces and NACs) to sample the nuclear trajectories at (hybrid) mean-field scaling. The electronic wavefunctions at each point are represented as variationally optimized linear combinations of the training states, which are themselves optimized in the training phase at selected nuclear geometries. This relies on a consistent and transferable representation of the many-body probability amplitudes, described in an atomic-local representation to facilitate this transferability of the low-energy probability amplitudes to different nuclear configurations. We discover a surprisingly broad spectrum of surface-hopping trajectories for this hydrogen chain, where the simple dimerization of all atomic pairs represent only a minority of the observed trajectories. While nuclear quantum effects are likely to contribute in this system and be an interesting direction for future work, the transitions are below the total initial energy and proceeds energetically downhill therefore unlikely to be a leading order effect. The nonadiabatic dynamics of this simple system could therefore represent an effective sandpit in the further development and comparison of hybrid classical-quantum molecular dynamics schemes as well as their dependence on underlying correlated electronic structure methodologies.


From a machine-learning perspective, this approach circumvents many of the traditional approximations and assumptions of local energy based decompositions for machine-learned force field parameterizations. The presence of an explicit many-body wavefunction and variational energy at each nuclear configuration ensures an inductive bias away from poorly represented regions of phase space. All desired observables are accessible within the same model, with a smoothly varying and physical electronic state at all times. The scope for accelerating the numerous situations where multiple sequential electronic structure calculations are required is clear -- from molecular dynamics to geometry or transition path optimization, vibrational spectroscopy and beyond.

Of course, many questions still remain. Chief amongst them is the question of how the number of training points required for a given accuracy scales with the nuclear phase space sampled. While this is likely to be system-dependent, it builds on the fundamental question of how transferable the low-energy physics is between these locally represented many-body states. While the nuclear phase space of the H$_{28}$ was relatively small ($\mathbb{R}^{14}$) given the one-dimensional and inversion-preserving configurations sampled due to the initial conditions, it was still large enough to make this a highly non-trivial interpolation from just 22 training points. This question of training set size must also necessarily consider the effect of more realistic basis set sizes. Currently, limitations still exist in obtaining the training data, and more approximate methods will need to be investigated, as well as circumventing the quartic scaling memory costs for the 2-tRDMs which are required. These are all critical questions for the long-term viability of this approach, which holds the promise for efficient acceleration of a wide number of correlated electronic structure methods.


\ifjcp\else
    \newpage
\fi

\section*{Acknowledgements}
We gratefully acknowledge support from the Air Force Office of Scientific Research under award number FA8655-22-1-7011 and the UK Materials and Molecular Modelling Hub for computational resources, which is partially funded by EPSRC (EP/T022213/1, EP/W032260/1 and EP/P020194/1). YR also acknowledges the support of EPSRC (EP/Y005090/1). 
\ifjcp
%
\else
    \bibliographystyle{achemso}
    \bibliography{ref.bib}
\fi










\end{document}